\documentclass[journal=jacsat,manuscript=article,layout=twocolumn]{achemso}

\newcommand*{\sometext}{Silicene and germanene derivatives constructed from periodic dumbbell units play a  crucial role in multilayers of these honeycomb structures.  Using first-principles calculations based on density functional theory, here we investigate the dumbbell formation mechanisms and energetics of Group IV atoms adsorbed on graphene, silicene, germanene and stanene monolayer honeycomb structures. The stabilities of the binding structures are further confirmed by performing ab-initio molecular dynamics calculations at elevated temperatures, except for stanene which is subject to structural instability upon the adsorption of adatoms. Depending on the row number of the adatoms and substrates we find three types of binding structures, which lead to significant changes in the electronic, magnetic, and optical properties of substrates. In particular, Si, Ge and Sn adatoms adsorbed on silicene and germanene form dumbbell structures. Furthermore, dumbbell structures occur not only on single layer, monatomic honeycomb structures, but also on their compounds like SiC and SiGe. We show that the energy barrier to the migration of a dumbbell structure is low due to the concerted action of atoms. This renders dumbbells rather mobile on substrates to construct new single and multilayer Si and Ge phases.

}

\let\oldmaketitle\maketitle
\let\maketitle\relax

\author{V. Ongun \" Oz\c celik}
\affiliation{UNAM-National Nanotechnology Research Center, Bilkent University, 06800 Ankara, Turkey}
\alsoaffiliation{Institute of Materials Science and Nanotechnology, Bilkent University, Ankara 06800, Turkey}
\email{ongunozcelik@bilkent.edu.tr}
\author{D. Kecik}
\affiliation{UNAM-National Nanotechnology Research Center, Bilkent University, 06800 Ankara, Turkey}
\alsoaffiliation{Institute of Materials Science and Nanotechnology, Bilkent University, Ankara 06800, Turkey}
\email{kecik@unam.bilkent.edu.tr}
\author{E. Durgun}
\affiliation{UNAM-National Nanotechnology Research Center, Bilkent University, 06800 Ankara, Turkey}
\alsoaffiliation{Institute of Materials Science and Nanotechnology, Bilkent University, Ankara 06800, Turkey}
\email{durgun@unam.bilkent.edu.tr }
\author{S. Ciraci}
\affiliation{Department of Physics, Bilkent University, Ankara 06800, Turkey}
\email{ciraci@fen.bilkent.edu.tr}

\title{Adsorption of Group-IV Elements on Graphene, Silicene, Germanene, Stanene: Dumbbell Formation}

\begin{document}

\twocolumn[
\begin{@twocolumnfalse}
\oldmaketitle
\begin{abstract}
\sometext
\end{abstract}
\end{@twocolumnfalse}
]

\section{Introduction}

Recent theoretical and experimental studies have proven that silicon,\cite{engin,seymur1,lelay2,lelay1,van2014free} germanium,\cite{hasan,lelay3,ongun_jpcl,rachel2014giant} compound semiconductors,\cite{hasan} $\alpha$-silica,\cite{silicatene} $\alpha$-tin,\cite{tin1,tin2,tin4,van2014two} transition metal oxides and dichalcogenides\cite{cantx2,tongay2012thermally,novotx2,wangtx2} can have stable, single layer honeycomb structures like graphene. However, in contrast to suspended graphene which can be easily exfoliated from 3D layered graphite; free-standing single layers of Si, Ge and Sn were not synthesized yet, since these elements do not exist as 3D layered bulk phase in nature. Therefore, it is a much accessible way (and the only possible way so far), to synthesize single layers of Si (i.e. silicene), Ge (i.e. germanene) and Sn (i.e. stanene) on suitable substrates like silver and gold. Under these circumstances, the growth of stable multilayers of silicene was recently achieved.\cite{lelay3,lelay4,24h} After the synthesis of thick layered silicene, the possibility of the layered bulk allotrope of silicon has been explored and stable bulk phases of Si have been predicted, which show a layered character and display electronic and optical properties different from those of the well-known cubic diamond structure.\cite{silicite} These results were further supported by the experimental data collected from multilayer silicene grown on Ag(111) substrate. \cite{lelay3,lelay4,24h}

\begin{figure*}
\includegraphics[width=16cm]{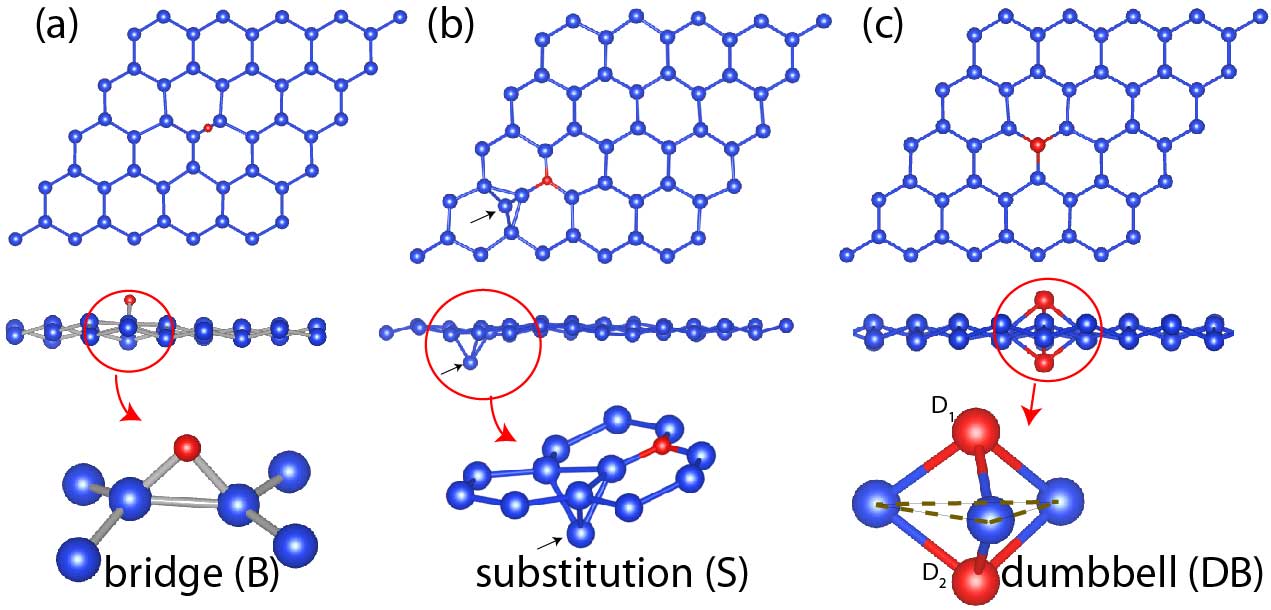}
\caption{Side and top views of various types of equilibrium atomic structures, which occur when a Group IV element (C, Si, Ge and Sn) is adsorbed on the single layer honeycomb structure constructed from Group IV elements (graphene, silicene and germanene). (a) A Group IV adatom adsorbed at the bridge site, i.e. B-site. (b) Substitution of silicene or germanene host atoms by the carbon adatom, i.e. S-site. The small arrows indicate the host Si atom that is pushed down after the substitution of C adatom. (c) Dumbbell structure (DB) constructed by Si, Ge and Sn adatoms adsorbed on silicene and germanene. D$_1$ and D$_2$ are dumbbell atoms; D$_1$ is the adatom, D$_2$ is the host atom pushed down by D$_1$. Adatom and single-layer substrate atoms are described by red and blue balls, respectively.}
\label{fig1}
\end{figure*}

Understanding the structure of layered silicene is of particular importance in a wide range of applications from electronics design to Li-ion storage for batteries.\cite{tritsaris2013adsorption}
Scanning tunneling microscopy (STM) measurements aiming at the understanding of the structure of multilayer silicene grown on Ag(111) substrate\cite{lelay3,lelay4} revealed that silicene layers have a $(\sqrt{3}\times \sqrt{3})$ supercell with honeycomb structure.  Earlier, theoretical studies found that the adsorption of a Si adatom on silicene is exothermic and results in a dumbbell structure (DB), where the Si adatom attached to the top side on silicene pushes the host Si atom down to form a cage.\cite{ongun1, ongun2, kaltsas,gimbert2014diverse} Recently, silicene derivatives constructed from periodic patterns of DB structures are found to be energetically more favorable.\cite{kok3} Additionally, stacking of these DB based silicene derivatives reproduced the structure data obtained from STM measurements of grown multilayers. These findings have pointed that the grown multilayer silicenes may, in fact, be constructed from the dumbbell based single layer phases of silicene.\cite{kok3} Furthermore, our recent letter\cite{ongun_jpcl} shows that  stable DB structures can occur not only on silicene, but also on germanene. Then the important question remains to be answered is whether DB based single layer phases can be common to all Group IV elements.

Motivated with the remarkable aspects of the DB structures, their insight on the layered allotropes and coverage depended phases revealed,\cite{ongun_jpcl} in this study we carried out an extensive analysis of the adsorption of the Group IV adatoms (C, Si, Ge and Sn) on the single layer honeycomb substrates constructed from these atoms (i.e. graphene, silicene, germanene, and stanene) amounting to 16 possible adatom+substrate combinations. Furthermore, we extended our analysis to include single-layers compounds, such as SiC and SiGe.  The question of whether the DB structure is common to all of these systems has been our starting point. Important findings of our study can be summarized as: (i) Three different types of equilibrium binding structures can occur when Group IV adatoms adsorb on graphene, silicene and germanene. These are specified as bridge bonding (B), substitutional (S), and dumbbell (DB). (ii) DB structure is not observed if the adatom or monatomic, single layer honeycomb substrate involves carbon atom. (iii) A structural instability is induced when one atom from Group IV elements is adsorbed on stanene, whereby the honeycomb structure is disrupted even at T=0 K. (iv) In silicene and germanene, C adatom substitutes host Si and Ge atoms. (v) Si, Ge and Sn form stable DB structures on silicene, as well as germanene with critical electronic, magnetic, and optical properties. (vi) In compound single layer honeycomb structure of silicon carbide, C adatom forms dumbbell structure; similar to Si and Ge adatoms adsorbed on single layer SiGe compound. (vii) We showed that the energy barrier for the migration of DBs is not high due to the concerted process of atoms at close proximity. This situation renders DB structures of Si, Ge and Sn mobile on silicene and germanene and paves the way to grow new single layer phases having different periodic patterns of the DBs. These phases, in turn, can offer alternatives to grow thin films or layered bulk structures and their compounds with diverse properties. (viii) Present results indicates that novel electronic, magnetic and optical properties can be achieved through the periodic coverage of graphene, silicene and germanene by Group IV adatoms.

\section{Method}
Our predictions were obtained from  first-principles pseudopotential calculations based on the spin-polarized density functional theory (DFT)\cite{dft1,dft2} within generalized gradient approximation (GGA) including van der Waals corrections.\cite{grimme} Projector-augmented wave potentials (PAW)\cite{blochl94} were used and the exchange-correlation potential was approximated with Perdew-Burke-Ernzerhof (PBE) functional.\cite{pbe} Adsorption of single adatom on various single layer honeycomb structures was simulated by using periodically repeating supercell method in terms of 5$\times$5 supercells comprising 50 host atoms and one adatom. The Brillouin zone (BZ) was sampled according to the Monkhorst-Pack scheme, where the convergence in energy as a function of the number of \textbf{k}-points was tested. The \textbf{k}-point sampling of (21$\times$21$\times$1) was found to be suitable for the BZ corresponding to the primitive unit cells of substrates. For larger supercells this sampling has been scaled accordingly. For the case of the 5$\times$5 cell used in this study, the \textbf{k}-point sampling was chosen as 5$\times$5$\times$1. Atomic positions were optimized using the conjugate gradient method, where the total energy and atomic forces were minimized. The energy convergence value between two consecutive steps was chosen as $10^{-5}$ eV. A maximum force of 0.002 eV/\AA~ was allowed on each atom. Numerical calculations were carried out using the VASP software.\cite{vasp} Since the band gaps are underestimated by standard-DFT methods, we also carried out calculations using the Heyd-Scuseria-Ernzerhof(HSE) hybrid functional,\cite{hse,heyd2003hybrid} which is constructed by mixing 25 \% of the Fock exchange with 75 \% of the PBE exchange and 100 \% of the PBE correlation energy. For the optical properties computed at the random phase approximation (RPA) level\cite{gajdos}, a (127$\times$127$\times$1) \textbf{k}-point grid and a total of 96 bands were undertaken for a proper description of the dielectric function.

The binding energy $E_b$, was calculated using the expression $E_{b} = E_{T} [A] + E_{T} [sub] - E_{T}[A+sub] $ in terms of the optimized total energies of adatom $E_{T}[A]$, of bare substrate (graphene, silicene, etc.) $ E_{T}[sub] $ and of adatom adsorbed on substrate $E_{T}[A+sub]$, all calculated in the same supercell. Positive values of $E_{b}$ indicate that the adsorption of the adatom is an exothermic process and is energetically favorable.

Further to conjugate gradient method, the stabilities of structures were tested by \emph{ab-initio} molecular dynamic (MD) calculations performed at finite temperatures. A Verlet algorithm was used to integrate Newton's equations of motion with time steps of $2fs$. We carried out MD calculations at temperatures $T=200K$,  $T=400K$, $T=600K$, $T=800K$ and $T=1500K$, each lasting $1ps$ and totaling to $5ps$ for each adatom+substrate system. To maintain the system in the desired constant temperature, the velocities of atoms were rescaled in each time step allowing a continuous increase or decrease of the kinetic energy.

\section{Equilibrium Structures and Energetics}
Adsorption of Group IV atoms on single layer graphene, silicene, germanene, and stanene are studied in terms of periodically repeating supercell method, where one adatom is adsorbed in each 5$\times$5 supercell. Since the adatom-adatom coupling between the 5$\times$5 supercells is minute, this system can be taken to mimic the single, isolated adatom and the local reconstruction thereof. The equilibrium binding structures and corresponding local reconstruction of C, Si, Ge and Sn adatoms on graphene, silicene, germanene, and stanene substrates have been explored by placing the adatoms at various sites and optimizing the atomic structures using conjugate gradient method. The optimized binding structures, namely B, S and DB are schematically described in  \ref{fig1}. Corresponding structural parameters, binding energy, magnetic and electronic structure are summarized in \ref{table1}. All of Group IV adatoms on stanene give rise to massive local reconstruction leading to the destruction of honeycomb structure in the 5$\times$5 supercell. This structural instability occurring even in the course of structure optimization using conjugate gradient method at $T=0K$ is critical, since it does not comprise artificial effects, like small unit cells enhancing stability etc. The instability followed by adatom adsorption implies that the single layer honeycomb structure of Sn is in a shallow minimum and is prone to structural deformations.

In particular, the carbon atom having electronic structure $1s^{2}2s^{2}2p^{6}$, behaves rather differently from the rest of the Group IV elements. For example, owing to relatively smaller C-C bond the $\pi$ - $\pi$ interaction stabilizes the planar structure of graphene attained by $sp^{2}$ bonding. Whereas single layer honeycomb structures of silicene, germanene and stanene are stabilized through buckling of atoms ensuring $sp^{3}$-like hybridization to compensate the weakening of $\pi$ - $\pi$ interaction. For the same reason, while carbon atoms can make stable monatomic chain structure (cumulene and polyyne),\cite{sefa, fan, chain, can1} suspended monatomic chain structure cannot be stable for the rest of Group IV elements.

\begin{figure*}
\includegraphics[width=16cm]{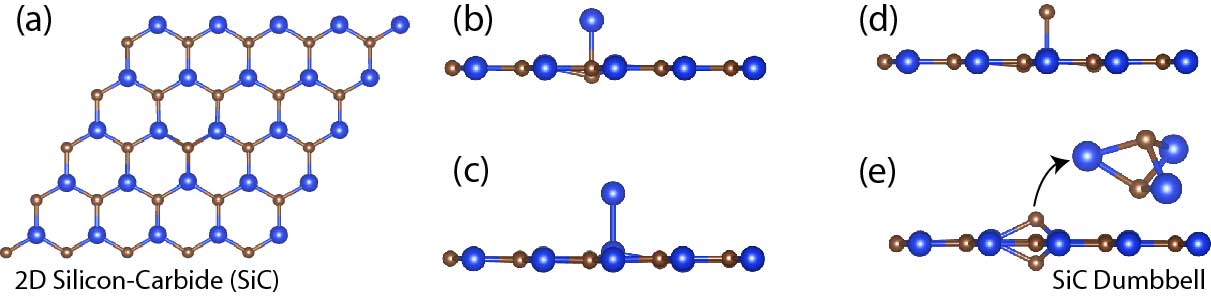}
\caption{ Equilibrium binding structures of C and Si adatoms on graphene like structure of SiC. (a) Top view of compound SiC displaying a planar honeycomb structure. (b) Si adatom on C host atom; (c) Si adatom on Si host atom; (d) C adatom on Si host atom are adsorbed at the head-on positions rather than forming a dumbbell structure. (e) C adatom on host C atom of SiC forming a DB structure. Si and C atoms are indicated by large blue and small brown balls, respectively.}
\label{sic}
\end{figure*}

Carbon adatom binds to graphene at B-site with a binding energy of $E_{b}$ $\simeq$ 1.7 eV as shown in \ref{fig1} (a). Even if C adatom is placed at the top site (which is 0.86 eV less favorable), it prefers to migrate back to the B-site.\cite{can1,canappl} As the C adatom coverage increases from 9$\times$9 to 2$\times$2, the binding energy can vary, but the B-site continues to be most favorable adsorption site. The binding energy was calculated to be 2.3 eV within the Local Density Approximation, which is known to yield over binding.\cite{can1,canappl} Upon the adsorption of C adatom at the B-site, while the underlying $C-C$ bond of graphene elongates and causes the weakening of $sp^2$ bonding, new bonds are formed between $2p$-orbitals of C adatom and graphene $\pi$-orbitals. The orbital composition and charge density of these new bonds were presented and the long range interaction between C adatoms were revealed.\cite{can1,canappl}. A chain structure situated perpendicular to the graphene can nucleate if additional C adatoms are placed at the close proximity of C adatom at B-site. On the other hand, carbon adatom adsorbed on silicene and germanene replaces the host Si or Ge atoms, respectively. These host atoms, which are removed from their position at the corner of hexagon, slightly dips below and moves towards the center of hexagon. They form three back bonds with three nearest atoms at the corners of hexagon as shown in \ref{fig1} (b). Substitutions of Si and Ge by C adatom or briefly S-type bindings are energetically favorable, since $Si-C$ and $Ge-C$ bonds are stronger than $Si-Si$ and $Ge-Ge$ bonds, respectively. Additionally, C adatom substituting Si or Ge host atoms becomes three-fold coordinated. Substitution energies are rather high and calculated as 5.89 eV and 5.02 eV for silicene and germanene, respectively. This situation suggests that one can generate new  derivatives from silicene and germanene through C substitution. On the other hand, the interactions of Si, Ge and Sn with graphene substrate and resulting binding structures are different. Since the length of a $C-C$ bond in graphene is smaller than $Si-C$, $Ge-C$ and $Sn-C$ bonds; Si, Ge and Sn adatoms adsorbed on graphene favor bridge bonds shown in  \ref{fig1} (a), where the top and hollow sites are 0.09 and 0.57 eV less favorable, respectively.\cite{akturk2010effects}

\begin{table*}
\caption{Adsorption of Group IV adatoms on single layer honeycomb structures of Group IV elements, namely graphene, silicene, and germanene in a 5$\times$5 supercell: Substrate (graphene, silicene or germanene); adatom; types of binding structure (B=bridge; S=substitution; DB=dumbbell); nearest adatom-substrate atom distance $d$ (values in parenthesis are D$_1$-D$_2$ distance in DB structures.); the lattice constant of the hexagonal unit cell of adatoms, $a$; magnetic moment per supercell ($\mu$); electronic state(numeral indicates the minimum band gap in eV, HSE band gaps are given in parenthesis); binding energy, $E_b$.}

\label{table1}
\begin{center}
\begin{tabular}{ccccccccc}
\hline  \hline
Substrate & Adatom & Structure & $d$(\AA)  & $a$(\AA) & $\mu$ ($\mu_B$) & Elec. State (eV) & $E_b$ (eV)\\
\hline
Graphene & C & B & 1.52 & 12.35 & 0.4 & metal & 1.670\\
Graphene & Si & B & 2.21 & 12.39 & 1.6 & metal & 0.799\\
Graphene & Ge & B & 2.41 & 12.41 & 1.6 & metal & 0.711\\
Graphene & Sn & B & 2.69 & 12.36 & 1.6 & metal & 0.808\\
Silicene & C & S & 1.96 & 19.24 & 2.0 & 0.26 (0.66) & 5.890\\
Silicene & Si & DB & 2.40 (2.70) & 19.28 & 2.0 & 0.08 (0.50) & 4.017\\
Silicene & Ge & DB & 2.51 (2.80) & 19.31 & 2.0 & 0.06 (0.47) & 3.612\\
Silicene & Sn & DB & 2.73 (2.97) & 19.21 & 2.0 & 0.09 (0.38) & 3.200\\
Germanene & C & S & 2.02 & 20.07 & 2.0 &  0.10 (0.42) & 5.026\\
Germanene & Si & DB & 2.46 (2.83) & 19.87 & 2.0 & 0.06 (0.39) & 3.200\\
Germanene & Ge & DB & 2.57 (2.93) & 19.89 & 2.0 & 0.06 (0.35) & 3.397\\
Germanene & Sn & DB & 2.76 (3.09) & 20.01 &  2.0 & 0.08 (0.30) & 3.080\\

\hline
\hline
\end{tabular}
\end{center}
\end{table*}

Energetically the most favorable binding structure of Si, Ge and Sn adatoms on silicene and germanene is the DB structure as described in the lower panel of  \ref{fig1}. The adatom (D$_1$) is first adsorbed on top of the host silicene or germanene. Subsequently it pushes down the host atom underneath (D$_2$) to form a DB consisting of D$_1$ and D$_2$ atoms. While D$_1$ lies above the substrate plane, D$_2$ is below and each one is bonded to three host atoms. The distance between D$_1$ and D$_2$ is relatively larger than the first nearest neighbor distance and consequently the bonding between them is weak. In this study, the DB atoms (D$_1$ and D$_2$) can be various combinations of  Group IV elements of the periodic table. However, owing to its weak stability, adsorption of Si, Ge and Sn adatoms on stanene do not form DB structures, rather they undergo local structural instability, followed by local destruction of single layer honeycomb structure upon adsorption of Si, Ge and Sn adatoms. Instability occurred not only in the course structure optimization of adatom+stanene system using structure optimization using conjugate gradient method, but also in \emph{ab-initio} MD calculations at low temperature.

DB structures form without any energy barrier once Si, Ge and Sn adatoms are placed on silicene or germanene.\cite{ongun1} The DB structure is of particular interest since specific periodic patterns of DBs on silicene or germanene can construct stable derivative structures which can have higher cohesive energy with different electronic and magnetic properties compared to parent silicene and germanene.\cite{ongun2,kok3,ongun_jpcl,silicite} These derivatives have shown to be crucial for the growth of multilayer silicene and germanene on Ag and Au substrates, respectively.\cite{kok3} In fact, stacking of these derivatives can make stable thin films\cite{lelay3,kok3} or 3D bulk layered structures, namely silicite and germanite, displaying rather different electronic and optical properties.\cite{silicite} Briefly, the synthesis of derivatives composed of silicene and germanene patterned by DBs of Si or Ge and their alloys pave the way towards nanostructures with physical properties different from their parent cubic diamond structure, cdSi or cdGe. Our results obtained from structure optimization indicate that the binding energy of a single DB on silicene and germanene is rather high and ranges between 4.01 eV and 3.08 eV.  Generally, while the binding energies decrease, $D_1$-$D_2$ distances increase as the row number of the Group IV adatom increases. We note that DB patterned phases of silicene and germanene display a side view of atoms; in particular D$_1$ and D$_2$ nematic orbitals are reminiscent of X-X bonds of transition metal dichalcogenides, MX$_2$.\cite{gimbert2014diverse} However, the FDS structure consisting of silicene having three DBs at the alternating corners of hexagon, which is, in fact, very similar to single layer MX$_2$, was found to be unstable.\cite{kok3}

It is known that single layer Group IV-IV compounds, like SiC and SiGe\cite{hasan} can be constructed. In SiC, Si and C atoms are located at the alternating corners of the hexagon to make a planar, graphene like structure with a 2.5 eV indirect band gap. Here, the binding structures of C and Si adatoms on C and Si host atoms of SiC are of interest. In spite of the fact that C adatom is adsorbed at B-site on graphene and substitutes Si host atom on silicene, C adatom forms a DB structure on SiC when placed on top of the host C atom. While the DB structure made of two C atoms is the second most energetic binding structure of graphene,\cite{akturk2010effects} DB of C atoms becomes the most energetic binding structure in SiC. On the other hand, C adatom on top of host Si, Si adatom on top of host C and Si adatom on top of host Si atom of SiC are bound at the top site (T). Even if the adatoms were displaced from the top site or were placed to B-site, they always moved to the equilibrium T site to minimize the total energy. In ~\ref{sic}, these four equilibrium binding structures, each with a magnetic moment of 2.0 $\mu_B$/per cell, are illustrated.

The SiGe honeycomb structure, where Si and Ge atoms are alternatingly located at the corner of hexagon, is another important compound we considered in this study. The buckling between adjacent Si and Ge is larger than that in silicene, but smaller than that in germanene. It is stable and has $\pi$-$\pi^*$ bands crossing linearly at the Fermi level, if small spin-orbit coupling is neglected. Four types of DBs, namely Si-Si, Ge-Si, Si-Ge and Ge-Ge DBs can be constructed, each having a magnetic ground state of 2.0 $\mu_b$/per cell; but attributing different physical properties to the SiGe substrate.

The crucial issue to address now is whether the adatom adsorbed substrates are stable. While optimized structure through conjugate gradient method provides evidence that the structure in hand is stable at T=0 K, this may correspond to a shallow minimum and hence the system may be destabilized at elevated temperatures. In fact, we found that the adsorption of all Group IV adatoms caused stanene to undergo a structural instability. Here we explored the stability of Group IV adatoms adsorbed on silicene and germanene at elevated temperature by performing \emph{ab-initio} MD calculations at finite temperatures, whose details are explained in the Methods section. Even at a temperature as high as $T=1500K$, the systems presented in \ref{table1} remained stable. We also note that similar MD calculations were carried out to test whether bare stanene by itself is stable at elevated temperatures. While free standing stanene maintained its structural stability for one picosecond at 400K, its structure is massively deformed during MD simulation at 600K. This explains why the structure of stanene is destroyed upon the adsorption of Group IV atoms treated in this study. It should be noted that stanene grown on substrate can attain stability and remain stable upon the adsorption of adatoms.

\section{Electronic Structures}
\begin{figure}
\includegraphics[width=8cm]{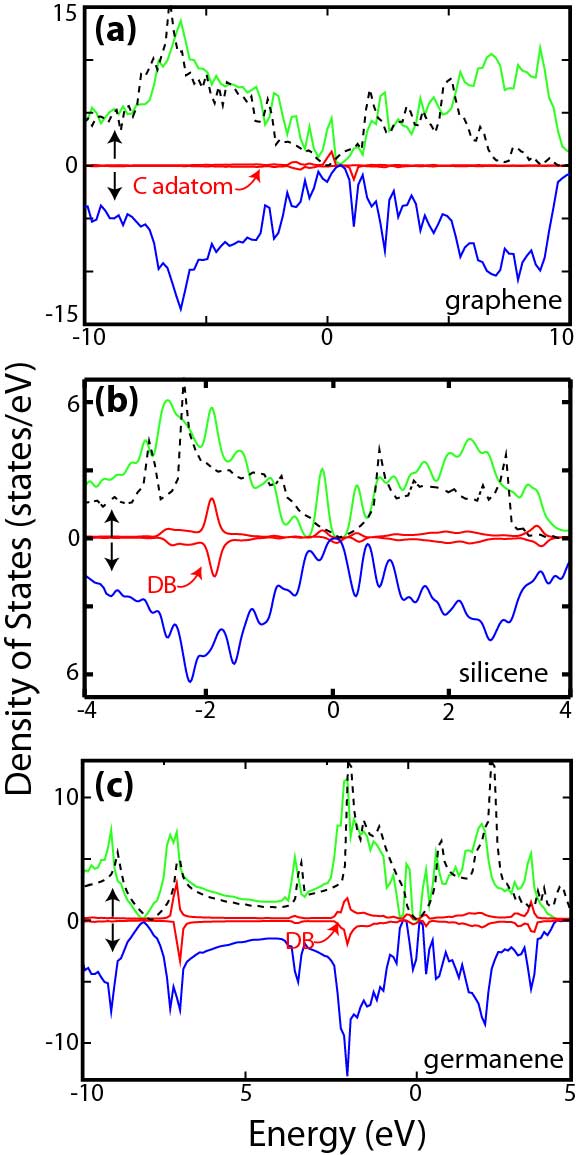}
\caption{Total density of states for (a) C on graphene (b) Si on silicene and (c) Ge on germanene. The partial DOS of the adatoms are indicated by red lines and multiplied by a factor of $2$ for better visualization. Spin up and spin down states are shown with green and blue lines, respectively. DOS of pure graphene, silicene and germanene are also indicated by the dashed curves in the plots.}
\label{figdos}
\end{figure}

\begin{figure*}
\includegraphics[width=14cm]{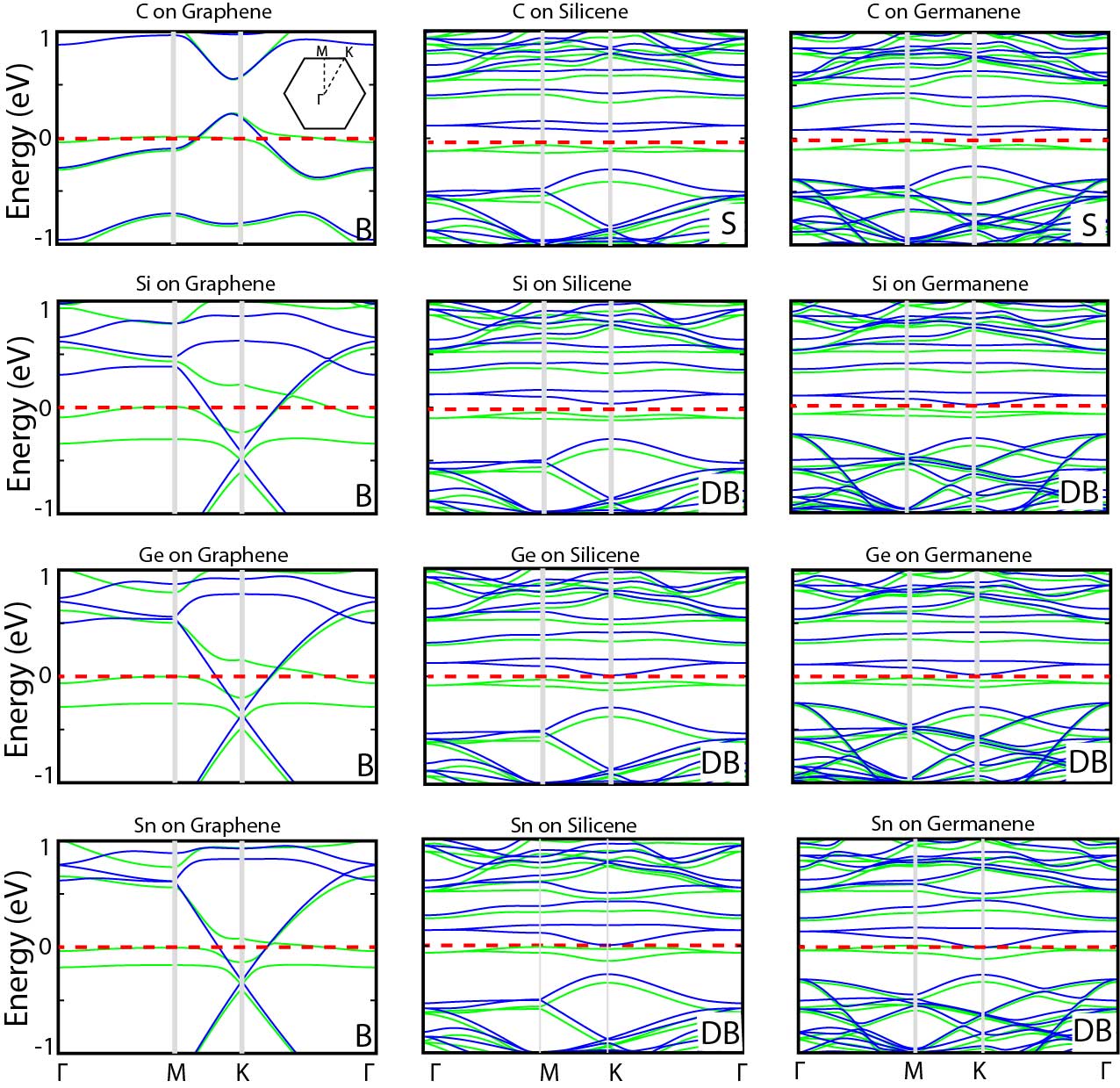}
\caption{Electronic band structures of monolayer graphene, germanene, and silicene with adatoms C, Si, Ge and Sn, which form a 5$\times$5 supercell or pattern. In the spin polarized systems, spin up and spin down bands are shown with green and blue lines. The zero of energy shown by red dashed line is set to the energy of highest occupied state. Binding structures (as B, S, DB) are indicated in each panel. Brillouin zone and symmetry directions are shown by inset.}
\label{fig2}
\end{figure*}

\begin{figure}
\includegraphics[width=8cm]{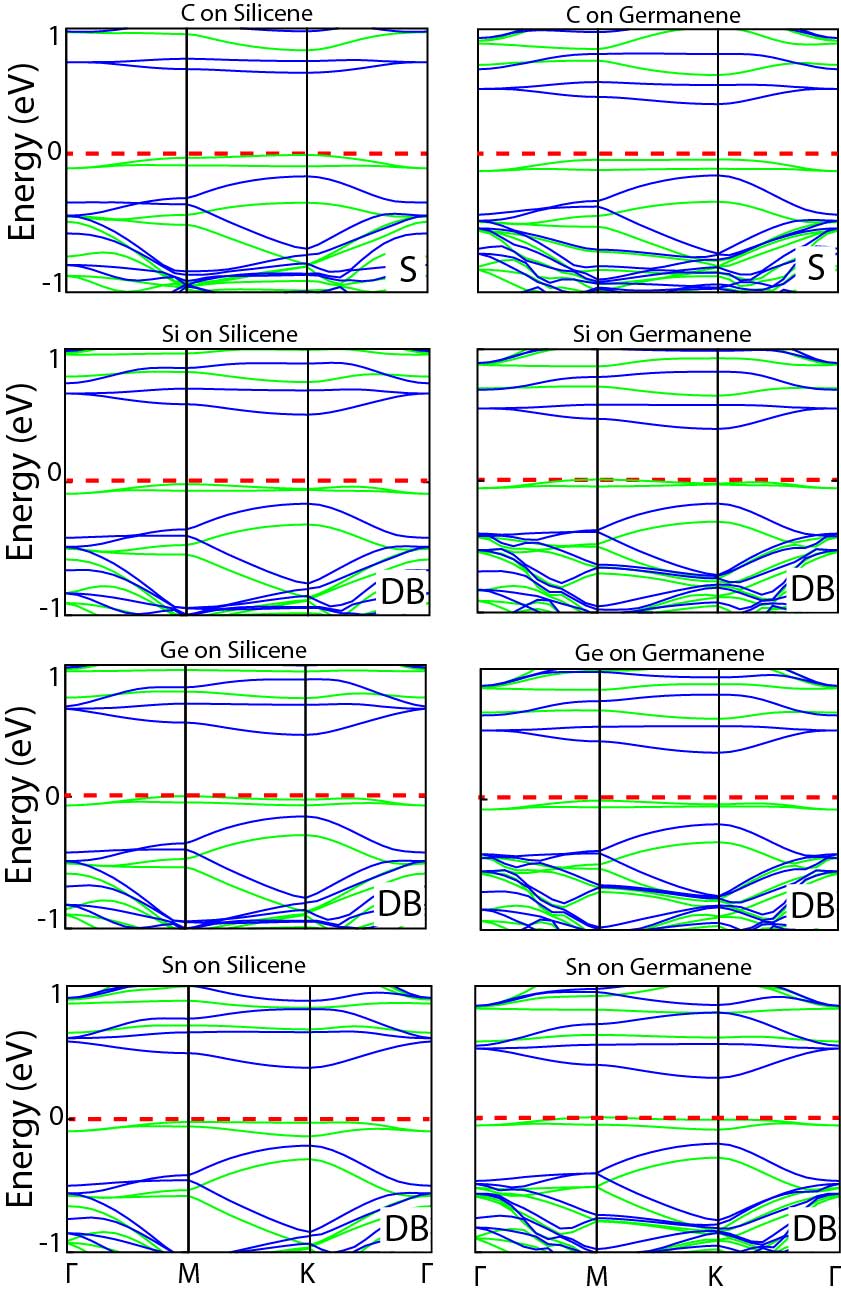}
\caption{Electronic band structures are calculated by using HSE for single layer silicene and germanene with adatoms C, Si, Ge and Sn, which form a 5$\times$5 supercell or pattern. In the spin polarized systems, spin up and spin down bands are shown with green and blue lines. The zero of energy shown by red dashed line is set to the energy of highest occupied state. Binding structures (as B, S, DB) are indicated in each panel.}
\label{hse}
\end{figure}

Group IV adatoms give rise to important changes also in the magnetic and electronic properties. In ~\ref{figdos} we presented the total density of states for C, Si and Ge adatoms adsorbed on graphene, silicene and germanene, respectively. The partial density of states of adatoms depicts the contribution of the adatom in the relevant energy range. For the sake of comparison, the density of states of corresponding bare substrate is also shown in each panel. While bare graphene, silicene and germanene all have nonmagnetic ground state, they attain magnetic ground state upon the adsorption of adatoms. The dominant effect of the adatom appears as sharp peaks near the band edges, which originate from the flat bands constructed from the mixing of orbitals of adatoms and host atoms at close proximity. In ~\ref{fig2}  we present the electronic band structures of the optimized structures for C, Si, Ge and Sn adatoms adsorbed also on graphene, silicene and germanene. In addition, we improve the band structures with HSE as shown in ~\ref{hse}. Similar to binding structure and energetics of adatom, the effect of the adatom on the electronic and magnetic structures are explored using a supercell model. If the spin-orbit coupling is neglected, bare graphene, silicene, as well as germanene are semimetals with conduction and valence bands crossing linearly at the Fermi level ($E_F$) carrying massless Dirac Fermion behavior.\cite{seymur1}  The localized (or resonance) states of single, isolated Group IV adatoms can occur below or above their $E_F$. However, within the periodically repeating 5$\times$5 supercell model with minute DB-DB coupling, these states are slightly broadened and form adatom bands. Therefore, the flat bands in ~\ref{fig2} are associated with the localized states due to adatoms. The  effect of the adatoms on the electronic and magnetic properties are summarized also in ~\ref{table1}. On silicene and germanene they lead similar electronic and magnetic states. DB formations on silicene and germanene in 5$\times$5 supercell periodicity result in ferromagnetic semiconductors with small band gaps of 0.06 - 0.12 eV within DFT-GGA. The magnetic moment per supercell is calculated to be 2.0 $\mu_B$. The energy difference between magnetic and non-magnetic state is rather small ($\sim$0.1 eV) for all cases indicating that it is a low temperature property. In what follows we present a comprehensive analysis of electronic band structures.

In ~\ref{fig2}, flat bands at the $E_F$ are associated with the states localized at carbon adatom at B-site forming bridge bonds with two nearest neighbor C atoms of graphene. These bands, which pin the $E_F$  are derived from $p_y$ and $p_z$ orbitals of carbon adatom at B-site. In the 5$\times$5 supercell, the crossing bands of bare graphene split and $E_F$ dips into the valence band to attribute metallic character. The net magnetic moment of each supercell is $\mu=$1.6$\mu_B$.  Si, Ge and Sn adatoms, which are also bonded to graphene at B-site give rise to metallic band structures as shown in ~\ref{fig2}. The $\pi^{*} - \pi$ bands of bare graphene, which cross linearly at the K-point of Brillouin zone dip 0.1-0.3 eV below the $E_F$ upon the adsorption of Si, Ge and Sn adatoms at B-site. It appears that the flat bands associated with the localized $p$-orbital states of Si, Ge and Sn adsorbate occur above the band crossing point and pin the $E_F$. In addition, each supercell attains a magnetic moment of 2 $\mu_{B}$. This is a significant effect, which makes non-magnetic graphene, silicene, and germanene spin-polarized.

The flat bands shown in ~\ref{fig2} near the band gap of the substitutional C adatom forming 5$\times$5 pattern in silicene and germanene are associated with the localized (or resonance) $p$-orbital states of the adatom as well as the $p$-orbital states of the host Si atom that is displaced from its regular position as indicated by the small arrow in ~\ref{fig1}(b). The effect of substitutional C atom is the splitting of the  $\pi^{*} - \pi$ bands crossing at the k-point and hence transforming the semi-metal bare silicene/germanene to semiconductor.  The lower valence bands are also affected from the substitutional C atom.

Upon the formation of Si-DB on silicene the spin-degeneracy of the bands are lifted as shown in  ~\ref{fig2} and narrow band gap opens between spin up and spin down bands. It should be noted that the electronic and magnetic properties of silicene patterned by Si-DB strongly depend on the size and symmetry of the DB pattern. Similar effects occur also with Ge-DB and Sn-DB forming 5$\times$5 pattern in silicene and germanene, except that the band gap between spin up and spin down bands is closed as the row number of adatom increases.

Considering the fact that the energy band gaps are underestimated by standard DFT methods, we repeat the calculations by using hybrid functionals,\cite{hse} except for graphene where all configurations are metallic.  According to those results all the systems are ferromagnetic semiconductor confirming what is obtained at DFT-GGA level with an expected increase in energy band gaps. The corrected band gaps are given in parenthesis in ~\ref{table1}. Interestingly,  energy band gaps decrease as the row number of the adatom as well as the substrate increases.

\begin{figure}
\includegraphics[width=8cm]{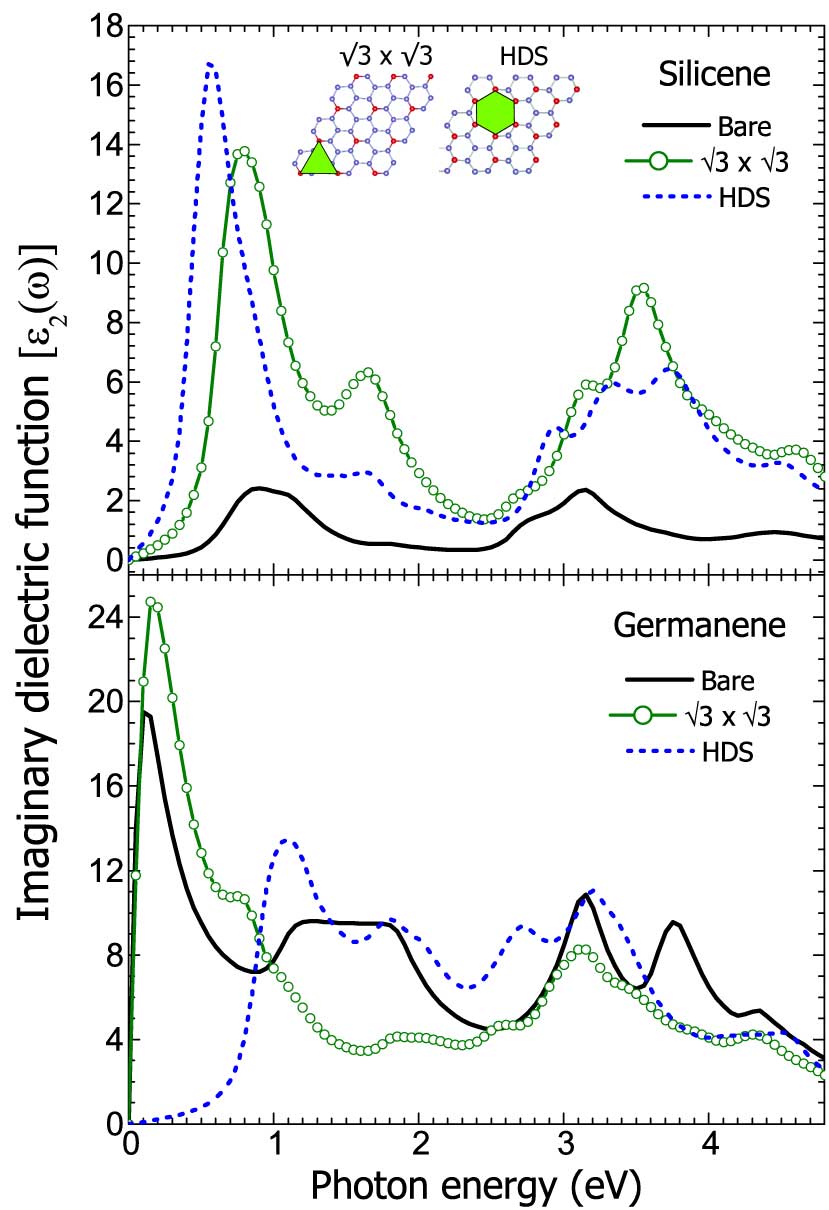}
\caption{The optical absorption spectra of bare (black curve), $(\sqrt{3}\times \sqrt{3})$ (green curve with dots) and HDS (dashed blue curve) structures of silicene (top) and germanene (bottom), respectively. Structures are shown in the inset plots.}
\label{figoptic}
\end{figure}

The optical properties of the two 2D honeycomb Group-IV crystals; silicene and germanene have been briefly investigated by calculating the frequency dependent complex dielectric function $\varepsilon_{i}(\omega)$ for normal incidence. The optical absorption is determined by the imaginary part of the dielectric function. The main peaks of absorption of both structures are related with the inter-band transitions that come into play. This finding is also supported in the works which study the infrared absorption spectra of silicene and germanene.\cite{Bechstedt1, Bechstedt2} In fact, both materials are known to be attractive candidates for nano-optoelectronic applications, since they display electronic and optical bandgaps which are within or in the vicinity of the visible part of the electromagnetic spectrum. Here the effects of Si and Ge adatoms are investigated by comparing the calculated optical properties before and after dumbbell formation. Two different coverages of dumbbells leading to $(\sqrt{3}\times \sqrt{3})$ dumbbell structure and hexagonal dumbbell structure (HDS) were investigated. \cite{kok3,ongun_jpcl} The imaginary dielectric function is displayed in ~\ref{figoptic} for both systems. Accordingly, bare silicene shows optical activity around 1 eV in the relatively early frequency regime, which extends to beyond 3 eV towards higher photon energies. As for the $(\sqrt{3}\times \sqrt{3})$ and HDS supercells of silicene, intense peaks of absorption are observed around 0.6 and 0.8 eV, respectively. Moreover, the second major peaks for both are concentrated in the range of 2.8-4.4 eV. $\varepsilon_{2}(\omega)$ of bare germanene, on the other hand, displays more similar features to its doped counterparts, when compared to silicene. An early strong absorption phenomenon takes place below 0.3 eV for both bare and $(\sqrt{3}\times \sqrt{3})$ forms of germanene. On the contrary, Ge-HDS shows a first peak around 1.1 eV. Some important preliminary results can be summarized as: (i) The intensities of the absorption peaks vary (i.e. become more distinguished) depending on the structure, when doped. (ii) The optically active region can be tuned by doping. Apparently, the periodic structure of DBs may introduce crucial effects on the optical absorption spectra, which may lead to certain potential applications in the visible range. Further investigation of the optical properties of DB-structures of silicene and germanene is considered as the topic of a future publication, which are aimed to be studied also at the level of many-body \emph{GW} corrections in order to introduce further accuracy to the peaks of absorption.

\begin{figure*}
\includegraphics[width=16cm]{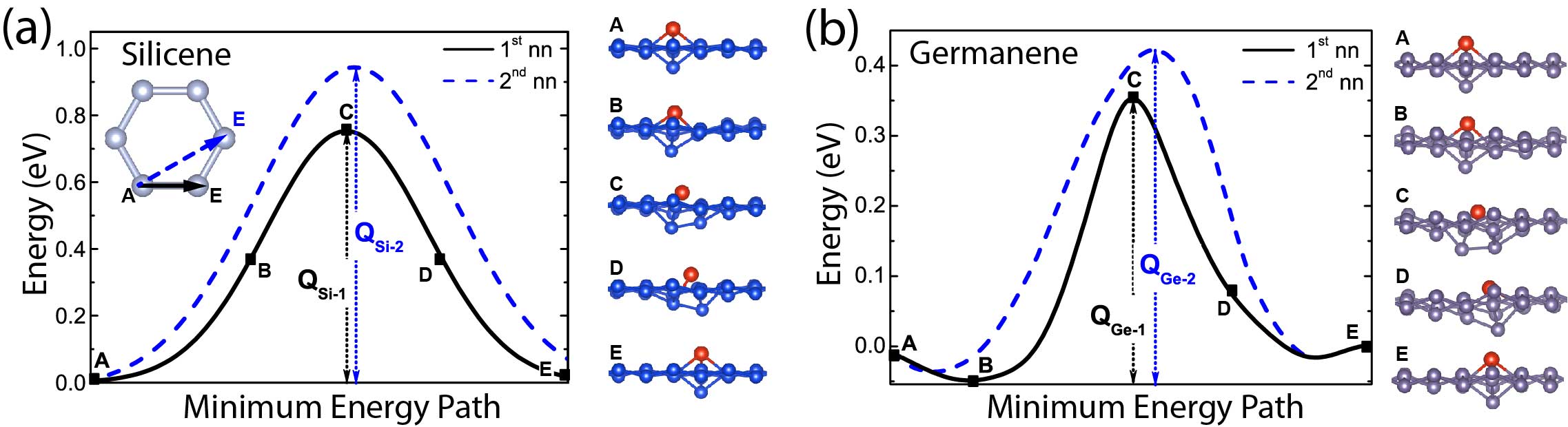}
\caption{(a) Migration of Si-DB on silicene and (b) Ge-DB on germanene. Energetics and the migration energy barrier ($Q$) of a single, isolated DB between the first and second nearest neighbors are shown by the inset in the left panel. The barriers are calculated using the nudged elastic band (NEB) method. Snapshots of the atomic configurations taken at stages A, B, C, D and E during the migration of DB to the first nearest neighbor site are also presented to illustrate the mechanism of diffusion.}
\label{fig3}
\end{figure*}

\section{Migration of Dumbbells on Silicene and Germanene}
As mentioned above, as soon as specific adatoms land on specific substrates, DBs can form through an exothermic process without an energy barrier. Moreover, DB-DB coupling is attractive until the first nearest neighbor separation.\cite{ongun2,ongun_jpcl} Consequently, as Si, Ge and Sn adatoms continue to land on silicene or germanene, firstly the domains of DBs form, eventually they join to cover the surface uniformly. In this respect, the migration or diffusion of DBs on silicene and germanene is crucial for the formation of various phases derived from silicene and germanene through their coverage by DBs in different concentration and symmetry. The energetics of the migration and the minimum energy barrier of single Si-DB on silicene and Ge-DB on germanene are investigated between the first and second nearest neighbors as shown in ~\ref{fig3} using nudged elastic band (NEB) method.\cite{neb}  The difference of the maximum and minimum total energy along the path corresponds to the energy barrier, $Q$. The energy barrier to the migration of D$_1$ of Si-DB along the straight path between the two first nearest neighbor atoms of silicene is $Q=$0.75 eV (minima are denoted by A and maxima by C in ~\ref{fig3}). Similar to the binding energies in \ref{table1}, the diffusion barrier of Ge-DB on germanene is lower than that of Si-DB on silicene. The barrier along the path between the second nearest neighbor is slightly higher. In a concerted action, as D$_1$ moves along the path, D$_2$ raises and eventually attains its original position at the corner of the buckled hexagon. On the contrary, the host atom at the first nearest neighbor site moves down as D$_1$ approaches and eventually the migration is completed with the construction of a new DB at the first nearest neighbor site. Snap-shots of atomic structure in the course of migration is also in shown ~\ref{fig3}. Low energy barrier to migration assures high mobility of DBs on silicene and germanene substrates needed for multilayer growth. The local minima denoted by B (which is energetically lower than A) for Ge-DB on germanene is considered to be due to a local unstable deformation appearing at that instantaneous atomic arrangement.

\section{Conclusions}
In this paper we investigated the binding of Group IV adatoms (C, Si, Ge and Sn) to single-layer, honeycomb structures of these atoms, namely graphene, silicene and germanene. The adsorption to stanene is not included since this structure is prone to instability upon the adsorption of any of the Group IV adatoms. Depending on the row number of a Group IV adatom, as well as substrates, we deduced three types of equilibrium binding structures. Isolated adatoms, as well as those forming periodically repeating supercells on graphene, silicene and germanene give rise to changes in electronic, magnetic and optical properties. Among the three types of binding structures, the dumbbell structure is of particular importance, since stable new phases of silicene and germanene can be derived from their periodic coverage with DBs. Dumbbell structures are also constructed on single layer, Group IV-IV compounds. The calculated energy barrier to the migration or diffusion of DBs on substrates is found to be low. This implies that DBs are rather mobile and cover the substrates as long as there is sufficient amount of incoming adatoms. By stacking these single layer phases one can grow thin film alloys and layered bulk structure of silicene and germanene.

\section{Acknowledgement}
The computational resources is provided by TUBITAK ULAKBIM, High Performance
and Grid Computing Center (TR-Grid e-Infrastructure). VO\"{O} and SC acknowledge  financial support
from the Academy of Sciences of Turkey(TUBA). ED acknowledges support from Bilim Akademisi - The Science Academy, Turkey under the BAGEP program. This work is partially supported by TUBITAK under the Project No. 113T050.

\bibliography{db_matrix_arxiv.bbl}

\end{document}